\documentclass[11pt,superscriptaddress,aps,prd,preprint]{revtex4}
\usepackage[dvips]{graphicx}
\usepackage{amsfonts}
\newcommand{\be}{\begin{equation}}
\newcommand{\en}{\end{equation}}
\newcommand{\ba}{\begin{eqnarray}}
\newcommand{\ea}{\end{eqnarray}}

\newcommand{\Slash}[1]{{#1}\!\!\!/}
\newcommand{\RM}[1]{\mathrm{#1}}

\begin{document}

\title{Dynamical Lorentz symmetry breaking in $3$D and charge fractionalization}
\author{B. Charneski}
\affiliation{Instituto de F\'{\i}sica, Universidade de S\~ao Paulo\\
Caixa Postal 66318, 05315-970, S\~ao Paulo, SP, Brazil}
\email{bruno,mgomes,tmariz,ajsilva@fma.if.usp.br}
\author{M. Gomes}
\affiliation{Instituto de F\'{\i}sica, Universidade de S\~ao Paulo\\
Caixa Postal 66318, 05315-970, S\~ao Paulo, SP, Brazil}
\email{bruno,mgomes,tmariz,ajsilva@fma.if.usp.br}
\author{T. Mariz}
\affiliation{Instituto de F\'{\i}sica, Universidade de S\~ao Paulo\\
Caixa Postal 66318, 05315-970, S\~ao Paulo, SP, Brazil}
\email{bruno,mgomes,tmariz,ajsilva@fma.if.usp.br}
\affiliation{UAST, Universidade Federal Rural de Pernambuco\\
Caixa Postal 063, 56900-000, Serra Talhada, PE, Brazil}
\email{tmariz@uast.ufrpe.br}
\author{J. R. Nascimento}
\affiliation{Departamento de F\'{\i}sica, Universidade Federal da Para\'{\i}ba\\
Caixa Postal 5008, 58051-970, Jo\~ao Pessoa, Para\'{\i}ba, Brazil}
\email{jroberto@fisica.ufpb.br}
\author{A. J. da Silva}
\affiliation{Instituto de F\'{\i}sica, Universidade de S\~ao Paulo\\
Caixa Postal 66318, 05315-970, S\~ao Paulo, SP, Brazil}
\email{bruno,mgomes,tmariz,ajsilva@fma.if.usp.br}

\begin{abstract}
We analyze the breaking of Lorentz invariance in a 3D model of fermion fields self-coupled through four-fermion interactions. The low-energy limit of the theory contains various sub-models which are similar to those used in the study of the graphene or in the description of irrational charge fractionalization.
\end{abstract}


\maketitle

\section{Introduction}

Four-fermion models have played an important role for the development of conceptual as well as practical aspects of quantum field theories. This observation is well illustrated by the works on the Thirring and Nambu-Jona-Lasinio models. More recently four-fermion interactions have been used in a variety of situations as in the study of phase diagrams relevant to the dynamics of high temperature cuprates \cite{hands} and in the study of the low-energy physics of graphene \cite{drut}. Four-fermion models have also been used in the investigation of the occurrence of Lorentz invariance breaking~\cite{Bjo}.

Although invariance under the continuous part of the Lorentz group, hereafter shortly referred as Lorentz invariance,  has been confirmed by many experiments, nowadays there is an increasing interest in the outcomes of highly sensitive experimental tests aimed to detect tiny violations of that symmetry in the standard model \cite{Kos}. Those deviations are expected to be low-energy manifestations of more fundamental theories unifying all known forces \cite{KosSam}.
To investigate this possibility, a standard model extension (SME) was constructed \cite{SME}. In that model,  through the Higgs mechanism, tensor fields acquire nonzero vacuum expectation values (VEV)  breaking Lorentz invariance spontaneously. The possible occurence of dynamical Lorentz symmetry breaking, a spontaneous breaking triggered by radiative corrections, in a self-interacting fermionic theory was considered in \cite{GomMar}.

The analysis above described were carried out in four dimensions, the investigations of Lorentz symmetry breakdown in 2+1 dimensions being  sparse \cite{Bel,Fer}, in  spite of, as mentioned before, its potential applications
to condensed matter systems. We should emphasize that differently of the $4D$ four-fermion model \cite{GomMar}, the $3D$ four-fermion model considered in this work is power counting renormalizable in the $1/N$ expansion \cite{Ros,GomRiv,GomMen}.

In this work, we will examine the possibility of causing a dynamical Lorentz symmetry breaking starting from self-interacting fermionic models which allow the induction of mixed Chern-Simons terms. As a mass-generating mechanism which preserves parity (P) and time-reversal (T), this doubled (or mixed) Chern-Simons term was first considered in \cite{JacPi}. Recently, it was shown that a low-energy effective theory for the fractional charge particles (portrayed by a honeycomb lattice \cite{Sem}, the lattice relevant to graphene \cite{Geim}) may be expressed in terms of a doubled Chern-Simons theory \cite{Ser,Ser2}. We also study a dynamical T (and P) symmetry breaking in which a conventional Chern-Simons term is induced. The breaking of the T symmetry has been recently used for a low-energy effective theory for the irrational fractional charge excitations \cite{Cha}.  

This paper is organized as follows. In Section \ref{model} we introduce the model and the notations. In Section \ref{EP} we fix the conditions for stable vacua  what gives rise to Lorentz and T (and P) violating VEV's. In Section \ref{LVMCS}, we prove that, due to Lorentz-violation, radiative corrections induce mixed Chern-Simons terms. Interesting enough, the low-energy effective Lagrangian obtained is similar to the one produced by integrating out the Dirac fermions on the graphene honeycomb lattice~\cite{Ser,Ser2}. This is discussed in the Section \ref{Gra} as well as the effective Lagrangian for systems with irrational fractional charge~\cite{Cha}. Section \ref{Con} contains a summary of results.

\section{The model}\label{model}

We begin by considering the fermionic model specified by the Lagrangian density
\begin{eqnarray}\label{tThirring}
{\cal L}_0 &=& \bar\psi(i\Slash{\partial}-m)\psi - \frac{G_S}2 (\bar\psi\psi)^2 - \frac{G_P}2 (\bar\psi i\gamma_5\psi)^2 - \frac{G_V}2 (\bar\psi\gamma_\mu\psi)^2 - \frac{G_{V_3}}2 (\bar\psi\gamma_3\psi)^2 \nonumber\\
&& - \frac{G_{\!A}}2 (\bar\psi\gamma_\mu\gamma_5\psi)^2 - \frac{G_{\!A_3}}2 (\bar\psi\gamma_3\gamma_5\psi)^2 - \frac{G_T}2 (\bar\psi\gamma_\mu\gamma_3\gamma_5\psi)^2 - \frac{G_{T_3}}2 (\bar\psi i\gamma_\mu\gamma_3\psi)^2,
\end{eqnarray}
where $\psi$ is a four-component spinor. For the three $4\times4$ $\gamma^\mu$ matrices ($\mu$ runs from 0 to 2) we choose the representation \cite{GomMen} 
$$
\gamma^0 = \left(
\begin{array}{cc}
\sigma^3 & 0\\
0 & -\sigma^3
\end{array}\right),\;
\gamma^1 = \left(
\begin{array}{cc}
i\sigma^1 & 0\\
0 & -i\sigma^1
\end{array}\right),\;
\gamma^2 = \left(
\begin{array}{cc}
i\sigma^2 & 0\\
0 & -i\sigma^2
\end{array}\right).
$$
The other two $4\times4$ matrices, $\gamma^3$ and $\gamma^5$, that anticommute with $\gamma^\mu$, are given by
$$
\gamma^3 = \left(
\begin{array}{cc}
0 & i\\
i & 0
\end{array}\right)\!,\;\mathrm{and}\;\mathrm{consequently}\;
\gamma^5 = i\gamma^0\gamma^1\gamma^2\gamma^3 =\left(
\begin{array}{cc}
0 & i\\
-i & 0
\end{array}\right)\!,
$$
so that $(\gamma^3)^2=-1$, $(\gamma^5)^2=1$, and $\RM{tr}(\gamma^\mu\gamma^\nu\gamma^\lambda\gamma^3\gamma^5)=-4i\epsilon^{\mu\nu\lambda}$. By using the identity $\gamma_5\sigma_{\mu\nu}=\epsilon_{\mu\nu\lambda}\gamma^3\gamma^\lambda$ (where $\sigma_{\mu\nu}=\frac i2[\gamma_\mu,\gamma_\nu]$) we observe that in Eq.~(\ref{tThirring}) we have included all the possible formally Lorentz invariants quadrilinear self-interaction terms. In order to rewrite the Lagrangian (\ref{tThirring}) in a more compact form, let us define the matrices
\be
\Gamma_{I}=\{1,\gamma_5,\gamma_\mu,\gamma_3,\gamma_\mu\gamma_5,\gamma_3\gamma_5,\gamma_\mu\gamma_3\gamma_5,\gamma_\mu\gamma_3\},
\en 
where the index $I$ assumes the values $S,P,V,V_3,A,A_3,T,T_3$ (with $\Gamma_S=1$, $\Gamma_P=\gamma_5$, and so on), so that we have
\begin{equation}
{\cal L}_0 = \bar\psi(i\Slash{\partial}-m)\psi - \frac{G_I}2 (\bar\psi\Gamma_I\psi)(\bar\psi\Gamma^I\psi).
\end{equation}

An arbitrary linear combination of quadrilinears $I=S,P,V_3,A,T,T_3$ does not induce any Chern-Simons terms. Furthermore, the quadrilinears $I=P,V,A_3,T,T_3$ induce the same Chern-Simons terms as the linear combinations of $I=V,V_3,A,A_3,T$. Thus, to avoid duplicity in the structure of the induced effective Lagrangian we will consider only the basic interaction given by a linear combination of
\be\label{Gamma}
\Gamma_I=\{\gamma_\mu,\gamma_3,\gamma_\mu\gamma_5,\gamma_3\gamma_5,\gamma_\mu\gamma_3\gamma_5\},
\en
i.e., $I$ takes only the values $V,V_3,A,A_3,T$. These quadrilinears induce three distinct mixed and one conventional Chern-Simons terms. Among the mixed Chern-Simons terms, two exhibit spontaneous Lorentz symmetry breaking (quadrilinears $I=V,V_3$ and $I=A,A_3$, with VEV $\langle \bar\psi\Gamma_A\psi \rangle\neq0$) and one does not (quadrilinears $I=V,T$). The quadrilinear $I=A$, with VEV $\langle \bar\psi\Gamma_{A_3}\psi \rangle\neq0$, induces the conventional Chern-Simons term which exhibits a spontaneous T (and P) symmetry breaking.

The bilinears $\bar\psi\Gamma^I\psi$ of these quadrilinears possess the discrete symmetries under the parity transformation defined as P: $\psi\to i\gamma^1\gamma^3\psi$, which implies in  
\be\label{P}
\begin{array}{l}
\bar\psi\psi(t,x,y) \to \bar\psi\psi(t,-x,y), \\
\bar\psi\gamma^5\psi(t,x,y) \to \bar\psi\gamma^5\psi(t,-x,y), \\
\bar\psi\gamma^\mu\psi(t,x,y) \to \bar\psi\{\gamma^0,-\gamma^1,\gamma^2\}\psi(t,-x,y) \\
\bar\psi\gamma^3\psi(t,x,y) \to -\bar\psi\gamma^3\psi(t,-x,y), \\
\bar\psi\gamma^\mu\gamma^5\psi(t,x,y) \to \bar\psi\{\gamma^0,-\gamma^1,\gamma^2\}\gamma^5\psi(t,-x,y), \\
\bar\psi\gamma^3\gamma^5\psi(t,x,y) \to -\bar\psi\gamma^3\gamma^5\psi(t,-x,y), \\
\bar\psi\gamma^\mu\gamma^3\gamma^5\psi(t,-x,y) \to \bar\psi\{\gamma^0,-\gamma^1,\gamma^2\}\gamma^3\gamma^5\psi(t,-x,y),
\end{array}
\en
the charge conjugation transformation defined as C: $\psi\to\bar\psi\gamma^0\gamma^1$, which implies in
\be\label{C}
\begin{array}{l}
\bar\psi\psi \to \bar\psi\psi, \\
\bar\psi\gamma^5\psi \to -\bar\psi\gamma^5\psi, \\
\bar\psi\gamma^\mu\psi \to -\bar\psi\gamma^\mu\psi, \\
\bar\psi\gamma^3\psi \to \bar\psi\gamma^3\psi, \\
\bar\psi\gamma^\mu\gamma^5\psi \to -\bar\psi\gamma^\mu\gamma^5\psi, \\
\bar\psi\gamma^3\gamma^5\psi \to \bar\psi\gamma^3\gamma^5\psi, \\
\bar\psi\gamma^\mu\gamma^3\gamma^5\psi \to \bar\psi\gamma^\mu\gamma^3\gamma^5\psi.
\end{array}
\en
The time-reversal transformation is the product of complex conjugation operator and an unitary operator. We choose $\psi\to\gamma^2\gamma^3\psi$, what implies   
\be\label{T}
\begin{array}{l}
\bar\psi\psi(t,x,y) \to \bar\psi\psi(-t,x,y), \\
\bar\psi\gamma^5\psi(t,x,y) \to -\bar\psi\gamma^5\psi(-t,x,y), \\
\bar\psi\gamma^\mu\psi(t,x,y) \to \bar\psi\{\gamma^0,-\gamma^i\}\psi(-t,x,y) \\
\bar\psi\gamma^3\psi(t,x,y) \to \bar\psi\gamma^3\psi(-t,x,y), \\
\bar\psi\gamma^\mu\gamma^5\psi(t,x,y) \to \bar\psi\{-\gamma^0,\gamma^i\}\gamma^5\psi(-t,x,y), \\
\bar\psi\gamma^3\gamma^5\psi(t,x,y) \to -\bar\psi\gamma^3\gamma^5\psi(-t,x,y), \\
\bar\psi\gamma^\mu\gamma^3\gamma^5\psi(t,-x,y) \to \bar\psi\{-\gamma^0,\gamma^i\}\gamma^3\gamma^5\psi(-t,x,y).
\end{array}
\en
The above transformations will be useful to analyze the induced Chern-Simons terms.

\section{Effective potential}\label{EP}

The self-interaction terms in Eq.~(\ref{tThirring}) may be eliminated through the introduction of the auxiliaries fields ${\cal A}_I=\{V_\mu,\Theta,A_\mu,\Phi,T_\mu\}$, so that the Lagrangian can be rewritten as
\begin{eqnarray}\label{Lag}
{\cal L} &=& {\cal L}_0 + \frac{g_I^2}{2}\left({\cal A}_I - \frac{e_I}{g_I^2} \bar\psi\Gamma_I\psi\right)^2 \nonumber\\
 &=& \frac{g_I^2}{2}{\cal A}_I{\cal A}^I + \bar\psi(i\Slash{\partial} - m - e_I{\cal A}_I\Gamma^I)\psi,
\end{eqnarray}
where $G_I=e_I^2/g_I^2$ and  $\Gamma_I$ is defined as in (\ref{Gamma}). By integrating over the fermion fields, the functional generator
\begin{equation}
Z(\bar \eta,\,\eta) = \int D{\cal A}_I D\psi D\bar\psi e^{i\int d^3x({\cal L}+\bar\eta\psi+\bar\psi\eta)}
\end{equation}
yields
\be\label{GF}
Z(\bar \eta,\,\eta) = \int D{\cal A}_I \exp\left[iS_\RM{eff}[{\cal A}] + i\int d^3x \left(\bar\eta\frac{1}{i\Slash{\partial} - m - e_I{\cal A}_I\Gamma^I}\eta \right) \right],
\en
where the effective action is given by
\begin{equation}\label{1}
S_\RM{eff}[{\cal A}] = \frac{g_I^2}{2}\int d^3x\, {\cal A}_I{\cal A}^I - i \RM{Tr} \ln(i\Slash{\partial} - m - e_I{\cal A}_I\Gamma^I)
\end{equation}
and $\RM{Tr}$ stands for the trace over Dirac matrices as well as the integration in momentum or coordinate variables. Thus, the effective potential turns out to be
\be\label{Vef}
V_\RM{eff} = - \frac{g_I^2}{2}{\cal A}_I{\cal A}^I + i\,\RM{tr}\int\frac{d^3p}{(2\pi)^3}\, \ln(\Slash{p} - m - e_I{\cal A}_I\Gamma^I),
\en
where the classical fields ${\cal A}^I$ are in a coordinate independent configuration. As we are interested in verifying the existence of nontrivial minima, we look for solutions of the equations
\begin{equation}\label{GapEq}
\frac{\partial V_\RM{eff}}{\partial{\cal A}_I}\Big|_{{\cal A}=b/e} =  - \frac{g_I^2}{e_I} b^I - i\,\Pi^I(m) = 0,\label{DVefDA} \\
\end{equation}
where $b_I=\{v_\mu,\theta,b_\mu,\phi,t_\mu\}$ are the minima and 
\begin{equation}\label{TadPole}
\Pi^I(m) = \RM{tr} \int\frac{d^3p}{(2\pi)^3} \frac{i}{\Slash{p} - m - b\cdot\Gamma}(-ie_I) \Gamma^I
\end{equation}
is the one-loop tadpole amplitude. From this point on we will use $b\cdot\Gamma = b_I\Gamma^I$.

\begin{figure}[h]
	\centering
		\includegraphics{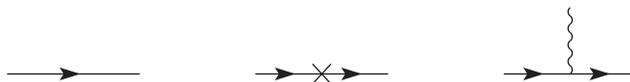}
  \caption{Feynman rules. Continuous and wave lines represent the fermion propagator and the auxiliary field, respectively. The cross indicates the $-ib_J\Gamma^J$ insertion in the fermion propagator and the trilinear vertex corresponds to $(-ie_I)\Gamma^I$}
	\label{fig1}
\end{figure}

\begin{figure}[h]
	\centering
		\includegraphics{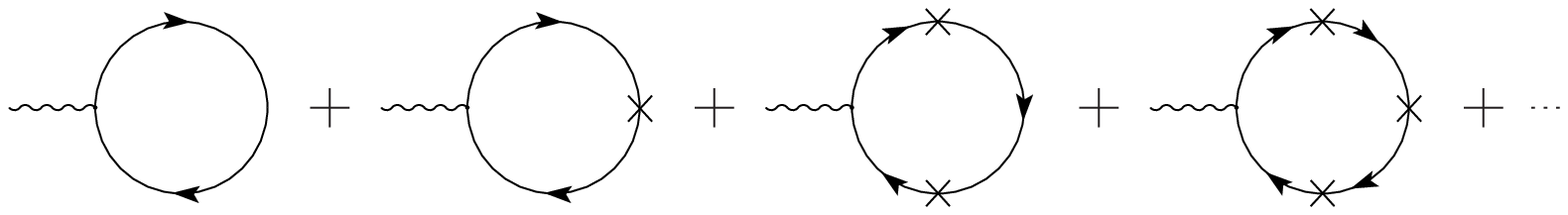}
	\caption{Contributions to the tadpole $\Pi^I$}
	\label{fig2}
\end{figure}

To evaluate the above integrals, we will follow the perturbative route where the free propagator is the usual one, $S(p)=i(\Slash{p}-m)^{-1}$, and the integrands are expanded in powers of $-ib\cdot\Gamma$. We proceed systematically by introducing a graphical representation so that, with the Feynman rules indicated in Fig.~\ref{fig1}, the contribution to $\Pi^I$ are shown in Fig.~\ref{fig2}. The first graph vanishes, but unlike what happens in four dimensions \cite{GomMar}, the third graph (with two insertions) as well as graphs with more than three insertions do not vanish. Using Pauli-Villars regularization with a regulator mass $M$ and renormalization spot $\mu$, the system of five equations with five unknowns implicit in Eq.~(\ref{GapEq}) becomes 
\begin{eqnarray}
\label{EqV}-\frac1{G_V} e_V\,v^\mu = 0,\;\;\; && \\
\label{EqTheta}\left(-\frac1{(G_{V_3})_R}+\frac{m-\mu}{\pi}+\frac{\theta^2}{2\pi m}+\frac{\phi^2}{\pi m}-\frac{t^2}{\pi m}+\frac{b^2}{2\pi m} + \cdots \right)e_{V_3}\,\theta + e_{V_3} \frac{b\cdot t}{\pi} + \cdots = 0,\;\;\; && \\
\label{EqB}\left(-\frac1{(G_A)_R}-\frac{m-\mu}{\pi}+\frac{\theta^2}{2\pi m}+\frac{\phi^2}{3\pi m}+\frac{t^2}{3\pi m}-\frac{b^2}{6\pi m} + \cdots\right)e_{\!A}b^\mu + \left(\frac{\theta}{\pi}-\frac{2b\cdot t}{3\pi m} + \cdots \right)e_{\!A}\,t^\mu = 0,\;\;\; && \\
\label{EqPhi}\left(-\frac1{(G_{A_3})_R}+\frac{2(m-\mu)}{\pi}+\frac{\theta^2}{\pi m}+\frac{b^2}{3\pi m} + \cdots \right)e_{\!A_3}\,\phi = 0,\;\;\; && \\
\label{EqT}\left(-\frac1{G_T}-\frac{\theta^2}{\pi m}+\frac{b^2}{3\pi m} + \cdots \right)e_T\,t^\mu + \left(\frac{\theta}{\pi}-\frac{2b\cdot t}{3\pi m} + \cdots \right)e_T\,b^\mu = 0,\;\;\; &&
\end{eqnarray}
where the ellipsis stand for terms with more than three insertions of $-ib\cdot\Gamma$ and  we have introduced the renormalized coupling constants
\begin{eqnarray}
\frac1{(G_{V_3})_R} &=& \frac1{G_{V_3}} + \frac{M+\mu}{\pi}, \\
\frac1{(G_A)_R} &=& \frac1{G_A} - \frac{M+\mu}{\pi}, \\
\frac1{(G_{A_3})_R} &=& \frac1{G_{A_3}} + \frac{2(M+\mu)}{\pi}.
\end{eqnarray}
Note that the $G_V$ and $G_T$ coupling constants are finite and then are not renormalized. Analysing these equations, we proceed as follows:

(i) From Eq.~(\ref{EqV}) we observe that $\langle V_\mu \rangle = v_\mu/e_V=0$. 

(ii) As can be verified, the VEV $\langle T_\mu \rangle = t_\mu/e_T$ does not generate any Chern-Simons term, so, without loss of generality we can take $t_\mu=0$. 

(iii)  With the choice $t_\mu=0$, the ellipsis in the second parenthesis of Eq.~(\ref{EqT}) contains only powers of $\theta$. Besides that, as in the situation in which dynamical Lorentz symmetry breaking occurs, $ b_\mu\neq 0 $, the expression inside the parenthesis must vanish. We enforce this last
requirement by choosing $ \theta=0 $ (as the determination of other nontrivial zeros is unfeasible). 

(iv) Adopting  the above restrictions, we conclude that the possible nonvanishing VEV's are $\langle A_\mu \rangle = b_\mu/e_{\!A}$ and $\langle \Phi \rangle = \phi/e_{A_3}$, and thus only the  Eqs. (\ref{EqB}) and (\ref{EqPhi}) need to be considered. 

(v) To have more insight on the gap equations mentioned in the previous item, we analyze them with more than three insertions. Up to seven insertions, we have
\begin{eqnarray}
\label{GapEq2}\left(-\frac1{(G_A)_R}-\frac{m-\mu}{\pi}+\frac{\phi^2}{3\pi m}-\frac{b^2}{6\pi m}-\frac{\phi^2b^2}{10\pi m^3}+\frac{b^4}{40\pi m^3}+\frac{3\phi^2b^4}{56\pi m^5}-\frac{b^6}{112\pi m^5}+\cdots\right)e_{\!A}b^\mu = 0,\;\; && \\
\label{GapEq3}\left(-\frac1{(G_{A_3})_R}+\frac{2(m-\mu)}{\pi}+\frac{b^2}{3\pi m}-\frac{b^4}{20\pi m^3}+\frac{b^6}{56\pi m^5}+\cdots\right)e_{\!A_3}\,\phi = 0.\;\; &&
\end{eqnarray}
As the general solution for these gap equations is not feasible, we fix $b_\mu$ to be lightlike, i.e., $b^2=0$, so that
\be\label{cond}
(G_{A_3})_R = \frac{\pi}{2(m-\mu)},
\en
and
\be\label{minphi}
\phi^2 = \frac{3\pi m}2\left(\frac1{(G_{A_3})_R} + \frac2{(G_A)_R} \right).
\en
Therefore, if the parenthesis in the above equation is positive, a nontrivial vacuum expectation value for the auxiliary scalar field will exist. In the following, we will see that due to the nonvanishing VEV's $\langle A_\mu \rangle$ and $\langle \Phi \rangle$, mixed and the conventional Chern-Simons terms will occur, respectively.

\section{Induced mixed and conventional Chern-Simons terms}\label{LVMCS}

Let us now study the fluctuations around the nontrivial minima of the potential. Writing ${\cal A}_I \to b_I/e_I + {\cal A}_I$, where $b_I=\{0,0,b_\mu,\phi,0\}$ and ${\cal A}_I=\{V_\mu,\Theta,A_\mu,\Phi,T_\mu\}$, so that now $\left\langle{\cal  A}_I \right\rangle = 0$, the generating functional (\ref{GF}) becomes
\be\label{Z}
Z(\bar \eta,\,\eta) = \int D{\cal A}_I \exp\left[iS_\RM{eff}[{\cal A},b] + i\int d^3x \left(\bar\eta\frac{1}{i\Slash{\partial}-m-b\cdot\Gamma-e_I{\cal A}_I\Gamma^I}\eta \right) \right],
\en
and the effective action is given by
\begin{equation}
S_\RM{eff}[{\cal A},b] = \int d^3x\left(\frac{g_I^2}{2}{\cal A}_I {\cal A}^I + \frac{g_I^2}{e_I}{\cal A}_I b^I + \frac{g_I^2}{2e_I^2}b_I b^I\right)-i \RM{Tr} \ln(i\Slash{\partial}-m-b\cdot\Gamma-e_I{\cal A}_I\Gamma^I).
\end{equation}
{In Eq.~(\ref{Z}), up to a field independent factor which may be absorbed in the normalization of the generating functional, we may replace $S_\RM{eff}[{\cal A},b]$ by the Pauli-Villars regulated expression
\begin{equation}
 \int d^3x\left(\frac{g_I^2}{2}{\cal A}_I {\cal A}^I + \frac{g_I^2}{e_I}{\cal A}_I b^I\right) + S^{\prime}_\RM{eff}[{\cal A},b],
\end{equation}
where
\be
 S^{\prime}_\RM{eff}[{\cal A},b] = \lim_{M\to\infty}\left(S^{\prime}_\RM{eff}[{\cal A},b,m] - S^{\prime}_\RM{eff}[{\cal A},b,M]\right)
\en
with 
\begin{equation}\label{series}
S^{\prime}_\RM{eff}[{\cal A},b,m] =\sum_{n=1}^\infty S^{(n)}[{\cal A},b,m],\qquad S^{(n)}[{\cal A},b,m]=i \RM{Tr} \frac1n\left[\frac i{i\Slash{\partial}-m-b\cdot\Gamma}(-ie_I){\cal A}_I\Gamma^I\right]^n
.
\end{equation}}
The formally divergent contributions in this formula are the tadpole, the self-energy, and the three point vertex function of the field ${\cal A}_I$.  The tadpole is given by 
\begin{eqnarray}
S_\RM{eff}^{(1)}[{\cal A},b,m] &=& i \RM{Tr} \frac i{i\Slash{\partial}-m-b\cdot\Gamma}(-ie_I){\cal A}_I\Gamma^I \nonumber\\
&=& i\int d^3x\, \Pi^I(m) {\cal A}_I,
\end{eqnarray}
where $\Pi^I=\{0,0,\Pi^A,\Pi^{A_3},0\}$ is part of the gap equations (\ref{GapEq}) evaluated in (\ref{GapEq2}) and (\ref{GapEq3}), with
\ba
\Pi^A &=& \left(\frac{M}{\pi}-\frac{m}{\pi}+\frac{\phi^2}{3\pi m}\right)ie_{\!A}b^\mu, \\
\Pi^{A_3} &=& \left(-\frac{2M}{\pi}+\frac{2m}{\pi}\right)ie_{A_3}\phi.
\ea

\begin{figure}[h]
	\centering
		\includegraphics{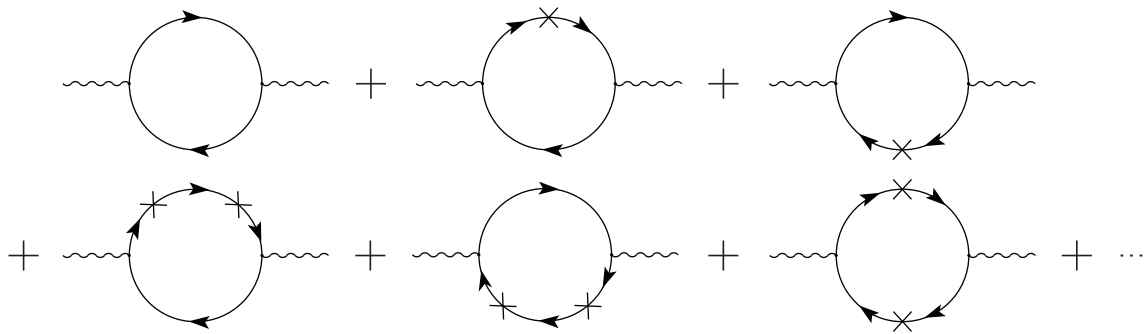}
  \caption{Contributions to the vacuum polarization $\Pi^{\mu\nu}$}
	\label{fig3}
\end{figure}

The self-energy term, which corresponds to $n=2$, yields
\begin{eqnarray}
S_\RM{eff}^{(2)}[{\cal A},b,m] &=& \frac i2 \RM{Tr} \frac i{i\Slash{\partial}-m-b\cdot\Gamma}(-ie_I){\cal A}_J\Gamma^J \frac i{i\Slash{\partial}-m-b\cdot\Gamma}(-ie_I){\cal A}_I\Gamma^I \nonumber\\
                      &=& \frac i2 \int d^3x\, {\cal A}_I\Pi^{IJ}(m){\cal A}_J,
\end{eqnarray}
with
\begin{eqnarray}\label{Pimunu}
\Pi^{IJ}(m)&=& \RM{tr} \int \frac{d^3p}{(2\pi)^4} \frac i{\,\Slash{p}-m-b\cdot\Gamma}(-ie_I){\Gamma^I} \frac {i}{\,\Slash{p}\,-\,i\Slash{\partial}-m-b\cdot\Gamma}(-ie_J)\Gamma^J.
\end{eqnarray}
By expanding in powers of $b\cdot\Gamma$, the above result can be expressed graphically as in Fig. \ref{fig3}. The second and third graphs are separately finite and furnish mixed and conventional Chern-Simons terms due to the nonvanishing VEV's $\langle A_\mu \rangle$ and $\langle \Phi \rangle$, whereas the first graph furnishes a mixed Chern-Simons term which does not break any symmetry. The divergent terms appear only from the first graph, where the Pauli-Villars regularization will also be used. By combining all the results, we get 
\ba
\label{VV}\Pi^{VV} &=& - \frac{ie_V^2}{6\pi m} (g^{\mu\nu}\Box-\partial^\mu \partial^\nu) - \frac{ie_V^2}{\pi m}b^\mu b^\nu, \\
\label{V3V3}\Pi^{V_3V_3} &=& \left(-\frac{M}{\pi}+\frac{m}{\pi}+\frac{\Box}{4\pi m}\right)ie_{V_3}^2 + \frac{ie_{V_3}^2\phi^2}{\pi m}, \\
\label{AA}\Pi^{AA} &=& g^{\mu\nu}\left(\frac{M}{\pi}-\frac{m}{\pi}+\frac{\phi^2}{3\pi m}\right)ie_{\!A}^2 - \frac{ie_{\!A}^2}{12\pi m} (g^{\mu\nu}\Box-\partial^\mu \partial^\nu) \nonumber\\ 
&& + \frac{ie_{\!A}^2}{12\pi m}\partial^\mu \partial^\nu - \frac{ie_{\!A}^2\phi}{3\pi m} \epsilon^{\mu\lambda\nu} \partial_\lambda - \frac{4ie_{\!A}^2}{3\pi m}b^\mu b^\nu + \frac{2ie_{\!A}^2\phi^2}{3\pi m}g^{\mu\nu}, \\
\label{A3A3}\Pi^{A_3A_3} &=& \left(-\frac{2M}{\pi}+\frac{2m}{\pi}+\frac{\Box}{6\pi m}\right)ie_{\!A_3}^2, \\
\label{TT}\Pi^{TT} &=& - \frac{ie_T^2}{6\pi m} (g^{\mu\nu}\Box-\partial^\mu \partial^\nu) - \frac{ie_T^2}{\pi m}b^\mu b^\nu,
\ea
for the diagonal part of $\Pi^{IJ}$, whereas for the off-diagonal we have
\ba
\Pi^{VV_3} &=& -\Pi^{V_3V} = \frac{ie_V\,e_{V_3}}{2\pi m}\epsilon^{\mu\nu\lambda}b_\nu \partial_\lambda, \\
\Pi^{V_3T} &=& \Pi^{TV_3} = -\frac{i}{\pi}e_{V_3}\,e_T\,b^\nu, \\
\Pi^{VT} &=& -\Pi^{TV} = \frac{i}{\pi}e_V\,e_T\,\epsilon^{\mu\lambda\nu}\partial_\lambda, \\
\Pi^{AA_3} &=& \frac{ie_{\!A}\,e_{\!A_3}}{6\pi m}\epsilon^{\mu\lambda\nu}b_\nu \partial_\lambda - \frac{2ie_{\!A}\,e_{\!A_3}}{3\pi m}\phi\,b^\mu, \\
\Pi^{A_3A} &=& \frac{ie_{\!A}\,e_{\!A_3}}{6\pi m}\epsilon^{\mu\lambda\nu}b_\mu \partial_\lambda - \frac{2ie_{\!A}\,e_{\!A_3}}{3\pi m}\phi\,b^\nu, 
\ea
where we disregarded terms of order of $\Box/m^2$, and higher. In these equations we see that the mixed Chern-Simons terms are in the off-diagonal part of $\Pi^{IJ}$ and the $\Pi^{AA}$ component has a conventional Chern-Simons term due to the VEV $\langle \Phi \rangle\neq0$. The other terms are kinetic, mass, and a gauge-fixing terms, and terms that complement the displaced bumblebee potential \cite{GomMar}.

\begin{figure}[h]
	\centering
		\includegraphics{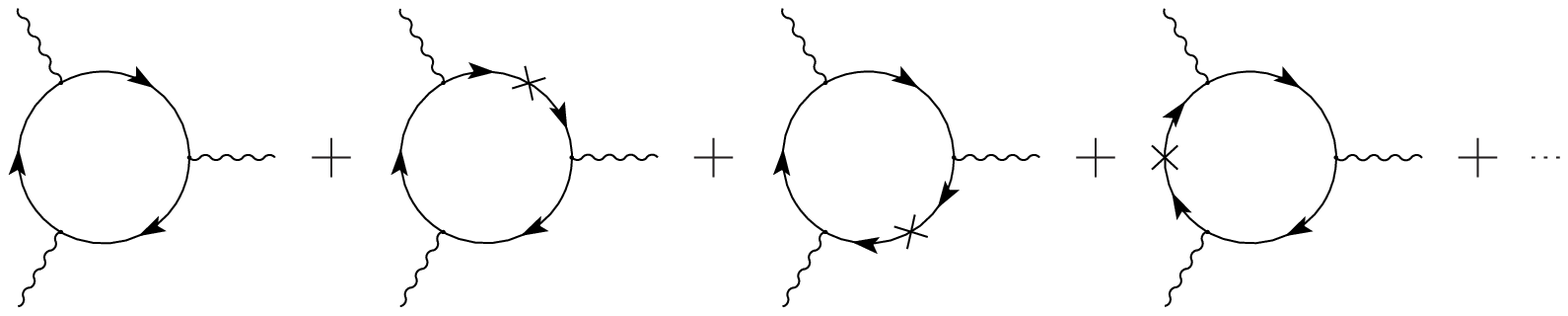}
	\caption{Contributions to the three-point $\Pi^{\mu\nu\rho}$}
	\label{fig4}
\end{figure}

In the third term of the series in Eq.~(\ref{series}), depending on which matrices are in the $(-ie_I)\Gamma^I$ vertex, the Furry theorem may be  applicable. Thus, we have
\be
S_\RM{eff}^{(3)}[{\cal A},b,m] =\frac i3 \int d^3x\,\Pi^{IJK}(m){\cal A}_I {\cal A}_J {\cal A}_K,
\en
where
\begin{eqnarray}\label{Pimnr}
\Pi^{IJK}(m) &=& \RM{tr} \int \frac{d^3p}{(2\pi)^4} \frac{i}{\,\Slash{p}-m-b\cdot\Gamma}(-ie_I) \Gamma^I \frac{i}{\,\Slash{p}\,-\,i\Slash{\partial}-m-b\cdot\Gamma}(-ie_J) \Gamma^J \nonumber\\ && \times \frac{i}{\,\Slash{p}\,-\,i\Slash{\partial}\,-\,i\Slash{\partial}'-m-b\cdot\Gamma}(-ie_K) \Gamma^K.
\end{eqnarray}
In the above formula the derivatives $\Slash{\partial}$ and $\Slash{\partial}'$ act on ${\cal A}_I$ and ${\cal A}_J$, respectively. The expansion of this expression in power series of $b\cdot\Gamma$ is given by the graph expansion of Fig.~\ref{fig4}. Evaluating the trace over the Dirac matrices, the first graph vanishes, whereas the leading terms (in the expansion in $\Box/m^2$) of the other graphs yield
\ba
\Pi^{AA_3A_3} &=& \Pi^{A_3AA_3} =\Pi^{A_3A_3A} = \frac{ie_{\!A}\,e_{\!A_3}^2}{3\pi m}b^\mu \\
\Pi^{AAA_3} &=& \Pi^{AA_3A} = \Pi^{A_3AA} = \frac{ie_{\!A}^2\,e_{\!A_3}}{3\pi m}\phi\, g^{\mu\nu} \\
\Pi^{AAA} &=& \frac{ie_{\!A}^3}{3\pi m}(g^{\mu\nu}b^\lambda + g^{\mu\lambda}b^\nu + g^{\nu\lambda}b^\mu).
\ea

In order to complete the displaced bumblebee potential we also consider the fourth term of the series in (\ref{series}), where the leading contribution is given by
\be
S^{(4)}_\RM{eff} = \int d^3x \left[\frac{e_{\!A}^4}{24\pi m}(A_\mu A^\mu)^2-\frac{e_{\!A}^2\,e_{\!A_3}^2}{12\pi m}\Phi^2 A_\mu A^\mu \right] + {\cal O}\left(\frac{\Box}{m^2}\right).
\en

{By performing the renormalization of the coupling constants as before, disregarding terms with $ n>4 $ and $ O(\Box/m^2) $, the results obtained so far allow us to write the effective Lagrangian as}
\be\label{lag}
{\cal L} = {\cal L}_\RM{K+M+GF} + {\cal L}_\RM{CS} + {\cal L}_{BB},
\en
where ${\cal L}_\RM{K+M+GF}$ represents the kinetic, mass and a gauge-fixing terms,
\ba
{\cal L}_\RM{K+M+GF} &=& \frac{g_V^2}2 V_\mu V^\mu - \frac{e_V^2}{24\pi m}F_{\mu\nu}^2(V) + \frac{e_{V_3}^2}{8\pi m}(\partial_\mu\Theta)^2 - \frac{e_{\!A}^2}{48\pi m}F_{\mu\nu}^2(A) + \frac{e_{\!A}^2}{24\pi m}(\partial_\mu A^\mu)^2 \nonumber\\
&& + \frac{e_{\!A_3}^2}{12\pi m}(\partial_\mu\Phi)^2 + \frac{g_T^2}2 T_\mu T^\mu - \frac{e_T^2}{24\pi m}F_{\mu\nu}^2(T),
\ea
${\cal L}_\RM{CS}$ the mixed and conventional Chern-Simons terms,
\ba
{\cal L}_\RM{CS} &=& - \frac{e_V\,e_{V_3}}{2\pi m}\Theta\,\epsilon^{\mu\lambda\nu}b_\mu\partial_\lambda V_\nu - \frac{e_V\,e_T}{\pi}\epsilon^{\mu\lambda\nu}V_\mu\partial_\lambda T_\nu + \frac{e_{V_3}\,e_T}{\pi}\Theta\,g^{\mu\nu}b_\mu T_\nu \nonumber\\
&& + \frac{e_{\!A}\,e_{\!A_3}}{6\pi m}\Phi\,\epsilon^{\mu\lambda\nu}b_\mu \partial_\lambda A_\nu + \frac{e_{\!A}^2}{6\pi m} \phi\,\epsilon^{\mu\lambda\nu} A_\mu\partial_\lambda A_\nu,
\ea
and ${\cal L}_{BB}$ the displaced bumblebee potential,
\ba
{\cal L}_\RM{BB} &=& \frac{e_{\!A}^4}{24\pi m} \left(A_\mu A^\mu - \frac4{e_{\!A}} A \cdot b\right)^2 + \frac{e_{\!A}^2e_{\!A_3}^2}{12\pi m}\left(\Phi-\frac2{e_{\!A_3}}\phi\right)^2A_\mu A^\mu \nonumber\\
&&+ \frac{e_{\!A}\,e_{\!A_3}^2}{3\pi m}\left(\Phi-\frac2{e_{\!A_3}}\phi\right)\Phi\,A \cdot b + \frac{e_V^2}{\pi m}(V\cdot b)^2 + \frac{e_V^T}{\pi m}(T\cdot b)^2,
\ea
with $F_{\mu\nu}(A)=\partial_\mu A_\nu - \partial_\nu A_\mu$. As we will see in the next section, the Eq.~(\ref{lag}), that from now on we call master Lagrangian, contains various models relevant to the study of condensed matter systems.

\section{Low-energy effective theory for fractional particles}\label{Gra}

An example of charge fractionalization in 2+1 dimensions is provided by the fractional quantum Hall effect, where the T symmetry is broken by a strong external magnetic field. Recently, it has been presented in the literature systems that exhibit charge fractionalization without the breaking of T symmetry \cite{Chan}, and it has been argued that the low-energy effective theory for these systems contains a mixed Chern-Simons term \cite{Ser,Ser2}. 
{By restricting our analysis to different sets of quadrilinears able to dynamically induce low-energy effective theories with mixed Chern-Simons terms, we investigate the three dimensional Lorentz symmetry violation in a ``relativistic'' condensed matter system (graphene)}.

Let us then analyze separately the models obtained by the dynamical Lorentz symmetry breaking, i.e. with $b_\mu\neq0$  starting with various sets of quadrilinears. Observing the master Lagrangian (\ref{lag}), we note that the first induced model is the sector of $V_\mu$ and $\Theta$ fields, obtained by starting with the quadrilinears $I=V,V_3,A$ only, yielding the Lagrangian
\ba\label{lagThetaV}
{\cal L}_{\Theta V} &=& \frac{e_{V_3}^2}{8\pi m}\partial_\mu\Theta\partial^\mu\Theta + \frac{g_V^2}2 V_\mu V^\mu -\frac{e_V^2}{24\pi m}F_{\mu\nu}^2(V) 
- \frac{e_{V_3}\,e_V}{2\pi m}\Theta\,\epsilon^{\mu\lambda\nu}b_\mu\partial_\lambda V_\nu \nonumber\\
&&- \frac{e_{\!A}^2}{48\pi m}F_{\mu\nu}^2(A) + \frac{e_{\!A}^2}{24\pi m}(\partial_\mu A^\mu)^2,
\ea
where $\left\langle A_\mu \right\rangle = b_\mu$ behaves like an external background field in the mixed Chern-Simons term. For simplicity, here and in what follows we are not writing the bumblebee terms apropriate for the specified choice of the starting quadrilinears.

 As the $A_\mu$ field in the above expression decouples, we will disregard the last two terms. Thus, by introducing $\vartheta(x)=x_\mu b^\mu\pm C$ where $C$ is a constant, we can rewrite the Lagrangian (\ref{lagThetaV}) up to a surface term as
\be\label{lagThetaV2}
{\cal L}_{\Theta V} = \frac{e_{V_3}^2}{8\pi m}\partial_\mu\Theta\partial^\mu\Theta + \frac{g_V^2}2 V_\mu V^\mu -\frac{e_V^2}{24\pi m}F_{\mu\nu}^2(V) + \frac{e_{V_3}\,e_V}{2\pi m}\,\vartheta\,\epsilon^{\mu\lambda\nu}\partial_\mu\Theta\,\partial_\lambda V_\nu.
\en

Another induced model, obtained by starting with the quadrilinears $I= V_{3}, A, T  $,  mixes the sectors of $\Theta$ and $T_\mu$ fields, but, as we can verify it does not contain a Chern-Simons-like term. Finally, let us consider the model which have a mixed Chern-Simons term and a conventional one, obtained by starting from the quadrilinears $I=A,A_3$, given by 
\ba\label{lagDeltaA}
{\cal L}_{\Phi A} &=& \frac{e_{\!A_3}^2}{12\pi m}\partial_\mu\Phi\partial^\mu\Phi - \frac{e_{\!A}^2}{48\pi m}F_{\mu\nu}^2(A) - \frac{e_{\!A_3}\,e_{\!A}}{6\pi m}\vartheta\,\epsilon^{\mu\lambda\nu}\partial_\mu\Phi\,\partial_\lambda A_\nu \nonumber\\
&&+ \frac{e_{\!A}^2}{6\pi m} \phi\,\epsilon^{\mu\lambda\nu} A_\mu\partial_\lambda A_\nu + \frac{e_{\!A}^2}{24\pi m}(\partial_\mu A^\mu)^2.
\ea
By making the choice $\phi=0$, $e_{\!A_3}=m$, and $\partial_\mu\Phi = B_\mu$ we obtain
\be\label{lagDeltaA2}
{\cal L}_{\Phi A} = -\frac{e_{\!A}^2}{48\pi m}F_{\mu\nu}^2(A) + \frac{m}{12\pi}B_\mu B^\mu - {{\frac{e_{\!A}\,\vartheta}{6\pi}\,\epsilon^{\mu\lambda\nu}B_\mu\partial_\lambda A_\nu}} + \frac{e_{\!A}^2}{24\pi m}(\partial_\mu A^\mu)^2,
\en
which, up to an adjustment of numerical constants and the presence of the Lorentz violation effect implicit in $\vartheta(x)$, is the low-energy effective theory for graphene discussed in \cite{Ser,Ser2}. As argued in \cite{Ser2},  the use of four component Dirac fermions is very important to construct a model describing a $ T $ preserving system. In the graphene the fermionic indices are associated to two Dirac points and two sub-lattices. Under time reversal, these two sub-lattices components transform between themselves and by radiative corrections generate the double Chern-Simons term.

An important difference between the models in the Eqs.~(\ref{lagThetaV2}) and (\ref{lagDeltaA2}) is the presence of the mass term for $V_\mu$ field in Eq.~(\ref{lagThetaV2}) which limits its effects to short distances.

Let us now choose $b_\mu=0$, i.e. no Lorentz symmetry breaking, and keep the other choices as before (in particular $ \theta=0$). We start from the quadrilinears $I=V_3,A,A_3$, so that the master Lagrangian becomes
\be\label{lagDeltaThetaA}
{\cal L}_{\Phi\Theta A} = \frac{e_{\!A_3}^2}{12\pi m}\partial_\mu\Phi\partial^\mu\Phi + \frac{e_{V_3}^2}{8\pi m}\partial_\mu\Theta\partial^\mu\Theta - \frac{e_{\!A}^2}{48\pi m}F_{\mu\nu}^2(A) + \frac{e_{\!A}^2}{6\pi m} \phi\,\epsilon^{\mu\lambda\nu} A_\mu\partial_\lambda A_\nu + \frac{e_{\!A}^2}{24\pi m}(\partial_\mu A^\mu)^2.
\en
Note that the Lagrangian (\ref{lagDeltaThetaA}) is similar to the one discussed in the Ref.~\cite{Cha} to  study the irrational charge fractionalization. However to describe charge fractionalization we would have to include from the start  the quadrilinears $I= S, P $ and take $ A_\mu $ and $ \Phi $ both equal to zero. 
More precisely, one would begin with the quadrilinears $I= S,P,V_3 $ so that the corresponding auxiliary fields  would be identified  with the components of the background parameters considered  in \cite{Cha}. As shown in that reference, the vacuum expectation value of the fermionic current, perturbatively
computed in a derivative expansion, corresponds to a topological identically conserved current which exhibits irrational charge (the charge becomes rational if the field $ A_\mu  $ is introduced). Time reversal can be spontaneously broken by adding the quadrilinear $ A_3 $ to the starting Lagrangian and requiring that the corresponding auxiliary field have a nonvanishing vacuum expectation value.

Besides these induced Lorentz and discrete symmetries violating models we have a model that does not break any symmetry, by considering the quadrilinears $I=V,T$, given by 
\be
{\cal L}_{VT} = \frac{g_V^2}2 V_\mu V^\mu - \frac{e_V^2}{24\pi m}F_{\mu\nu}^2(V) + \frac{g_T^2}2 T_\mu T^\mu - \frac{e_T^2}{24\pi m}F_{\mu\nu}^2(T) - \frac{e_V\,e_T}{\pi}\epsilon^{\mu\lambda\nu}V_\mu\partial_\lambda T_\nu.
\en
Actually, the last term in the above expression is absent if the Pauli-Villars regularization is used, nevertheless it may be present in the another regularization scheme. 

By using the P, C, and T transformation properties of the bilinears associated to the fields ${\cal A}_I$, Eqs.~(\ref{P}), (\ref{C}) and (\ref{T}), respectively, we constructed the Table 1 for the radiatively induced Chern-Simons terms.

\medskip
\centerline{\begin{tabular}{||c|c|c|c||c||}
\hline
$\mathcal{L}_{CS}$ & C & P & T & CPT \\ \hline 
$\vartheta\,\epsilon^{\mu\lambda\nu}\partial_\mu\Theta\,\partial_\lambda V_\nu$ & $-$ & $+$ & $+$ & $-$ \\ \hline
$\vartheta\,\epsilon^{\mu\lambda\nu}B_\mu\partial_\lambda A_\nu$ & $-$ & $+$ & $+$ & $-$ \\ \hline
$\phi\,\epsilon^{\mu\lambda\nu} A_\mu\partial_\lambda A_\nu$ & $+$ & $-$ & $-$ & $+$ \\ \hline
$\epsilon^{\mu\lambda\nu}V_\mu\partial_\lambda T_\nu$ & $+$ & $+$ & $+$ & $+$ \\ \hline
\end{tabular}}
\medskip

\centerline{\small Table 1: Discrete-symmetry properties for the Chern-Simons terms.}
\smallskip 

\section{Summary}\label{Con}

In this work we studied the breaking of Lorentz symmetry in a 3D four-fermion theory with especial emphasis on those interactions which allow for the generation of Chern-Simons terms. In the low-energy regime, the model contains various sub-models with important connections to the
condensed matter systems. If the breaking is triggered by a non-vanishing VEV of the composite field $\bar\psi\gamma_\mu\gamma_5\psi$ 
a mixed Chern-Simons term is generated. As discussed elsewhere, this term is important for a time-reversal preserving description of the dynamics of the graphene. It is interesting to mention that the Lagrangian (\ref{lagDeltaA}) is the dimensional reduction of a four dimensional Lorentz-violating Maxwell-Chern-Simons model \cite{Bel}. We also considered the dynamical break of the time-reversal symmetry, but preserving the Lorentz invariance; in this setting  an effective Lagrangian similar to the one used in the study of the irrational charge fractionalization was generated.

{\bf Acknowledgements.} This work was partially supported by Funda\c{c}\~{a}o de Amparo \`{a} Pesquisa do Estado de S\~{a}o Paulo (FAPESP) and Conselho Nacional de Desenvolvimento Cient\'{\i}fico e Tecnol\'{o}gico (CNPq). The work of T.~M. has been supported by FAPESP, project 06/06531-4.

\end{document}